\documentclass[letterpaper,12pt]{revtex4-1}   
\usepackage{epsfig}

\begin{document}

\title{Photonic lattices for astronomical interferometry}
\author{S. Minardi}
\address{ Institute of Applied Physics, Friedrich-Schiller University Jena, Max-Wien-Platz 1, 07743 Jena, Germany\\
e-mail: stefano@stefanominardi.eu}



\label{firstpage}

\begin{abstract}
Regular two-dimensional lattices of evanescently coupled waveguides may provide in the near future photonic components capable of combining interferometrically and simultaneously a large number of telescopes, thus easing the imaging capabilities of optical interferometers. In this paper, the theoretical modeling of the so-called Discrete Beam Combiners (DBC) is described and compared to the conventional model used for photonic beam combiners for astronomical interferometry. The performance of DBCs as compared to an ideal ABCD beam combiner is discussed and applications to astronomical instrumentation analyzed. 
 
\end{abstract}

\maketitle

\section{Introduction}

Long baseline, optical interferometry is the only sustainable option to achieve ultra high resolution imaging in astronomy \citep{Quirrenbach}. The retrieval of accurate interferometric images requires however a dense sampling of the spatial coherence function (complex visibility) of the light emitted by an astronomical target \citep{Haniff}, a task which can be eased by the simultaneous combination of light collected by several telescopes.
To this end, integrated optics can provide miniaturized multiple-telescope combiners which can be included in very compact  assemblies, while offering highly stable an repeatable visibility measurements  \citep{LETI1, LETI2}.
State-of-the-art photonic beam combiners can deliver simultaneous combination of up to four telescopes and provide visibility measurements over 6 interferometric baselines \cite{Benisty}. 
The modular and compact design of such components has significantly reduced the complexity of  instrumental design and relative calibration procedures, with considerable reduction of the time required to commissioning an instrument \citep{CLEO2011LAOG}. 
Existing photonic beam combiners are fabricated with silica-on-silicon technology, which constrains the waveguides and couplers to a plane. This represents a significant complication of the design of components allowing the combination of a significantly larger number of telescopes, due to the necessary presence of cross-overs and increased sensitivity to fabrication defects\citep{8combiner}. 
In this respect, three dimensional (3D) photonic components, such as those fabricated by direct laser-writing technique \citep{Itoh}, may provide a significant simplification of the design of interferometric beam combiners through off-plane fiber connections/couplers 
\citep{Thomson09,Rodenas12}.
The degree of simplification in design and fabrication of 'on chip' astronomical interferometers can indeed be dramatic, as proved by the recent proposal by \citet{olDBC} of using regular 2D-arrays of evanescently coupled waveguides (i.e. photonic lattices\citep{Christodoulides}) rather than a traditional cascade of couplers arranged in a 3D environment. The so called Discrete Beam Combiners (DBC), besides having a very simple design, have the potential of improving the scalability of integrated beam combiners to telescopic arrays of arbitrary size, while providing a slightly higher sensitivity respect to the ABCD \citep{Shao} or pupil remapping schemes \citep{Perrin}. Moreover, the possibility to fabricate DBCs with direct laser writing allows the possibility of cost effective prototyping on a range of materials suitable for virtually any optical wavelength from UV to MIR.

Aim of this paper is to provide an extended theoretical background of DBCs and a first analysis of the performance of the DBC scheme as compared to idealized beam combiners. In the first section the conventional theoretical modeling of photonic combiners (Visibility to PhotoMetry, V2PM, \cite{Lacour}) is recalled and linked to a formalism specifically developed for the analysis of the DBCs ($\alpha$-matrix approach).
In Section 3, after a short recall of the DBC concept and the theory of evanescent coupling of optical waveguides, the method to derive the $\alpha$-matrix for square arrays of waveguides is derived. In Section 4, the performance of ABCD combiners is compared to the DBC combining 3 to 6 telescopes. Section 5 concludes the paper indicating application perspectives of DBCs in astronomical interferometry. 

\section{Theory of interferometric beam combination}
\subsection{The V2PM approach}
Interferometric beam combiners are used to encode phase variations into intensity variations, suitable for the retrieval of the coherence properties of the combined optical fields. Ideally the output port of a beam combiner provides the interference signal of just one pair of the input fields combined by the device $N$ input complex fields  $\{A_{\mathrm{n}}\}$:
\begin{eqnarray}
I&=&\langle  A_1 A_1^* \rangle +\langle  A_2 A_2^* \rangle+2\Re\left[ \langle  A_1 A_2^* \rangle\right]\\
&=&\langle  A_1 A_1^* \rangle +\langle  A_2 A_2^* \rangle+2\left|  A_1 A_2^* \right|\gamma\cos(\phi_{12}+\phi_\gamma)
\end{eqnarray}
Here, $\langle\cdot\rangle$ indicates the time averaging, $\Re$ is the real part, $\phi_{12}$ is the average phase difference between the fields and $\gamma$ and $\phi_\gamma$ are the amplitude and phase of the complex visibility of the interfering fields.
More generally, the $M$ measurables $\{I_m\}$ of an interferometric beam combiner are the real part of time averages of all possible products of $N$ input complex fields  $\{A_{\mathrm{n}}\}$ with their conjugates (mutual coherences). This can be written in a compact form using the V2PM (visibility to pixel matrix) formalism \citep{Tatulli,Lacour}:
\begin{equation}
I_m=\Re \lbrace\sum_{n=1}^{L=\frac{N(N+1)}{2}} \lbrace V2PM \rbrace_{\mathrm{mn}} V_n\rbrace.
\end{equation}
Here, V2PM is a MxL complex-valued matrix, and the components of the vector $\mathbf{ V}$ are all possible time averaged products of the input fields with their conjugates:
\begin{equation}
\mathbf{ V}=
\left(
\begin{array}{c}
\langle  A_1 A_1^* \rangle\\\vdots \\ \langle A_NA_N^*\rangle\\ \langle A_1A_2^*\rangle\\ \vdots\\\langle A_1A_N^*\rangle\\\langle A_2A_3^*\rangle\\\vdots\\\langle A_{N-1}A_N^*\rangle
\end{array}
\right)
\end{equation}
The elements of the V2PM matrix are calculated from the complex transmission coefficients of the input fields to the $M$ outputs of the 
beam combiner. In existing planar photonic beam combiners exploiting the ABCD method \citep{Shao, Benisty}, the matrix is formed by 4xL blocks, as apparent from Fig. 1 \citep{Lacour}. The coherences are then usually obtained in the least square sense by  pseudo-inversion methods. In a real beam combiner, the fabrication defects give raise to deviations from the ideal matrix of Fig. 1 which may take the form of \textit{i)} amplitude variations of the non-zero elements, \textit{ii)} deviation from the ideal phase of the complex matrix elements, and \textit{iii)} finite amplitude and phase of the zero elements (cross-talk). 

\begin{figure}
\centering
${\left(
\begin{array}{c c c c c c}
 1 & 1 & 0 &  2 & 0 &  0 \\
1 & 1 & 0   &  2 e^{i\pi/2} & 0  & 0 \\
1 & 1 & 0  &  - 2                 & 0   & 0\\
1 & 1 & 0  &   2 e^{i3\pi/2}  & 0 & 0 \\
1 & 0 & 1 & 0 & 2 &  0 \\
1 & 0 & 1 & 0 &  2 e^{i\pi/2} & 0 \\
1 & 0 & 1 & 0  &  - 2               & 0\\
1 & 0 & 1 & 0  &   2 e^{i3\pi/2} & 0\\ 
0 & 1 & 1 & 0 &  0 & 2 \\
0 & 1 & 1 & 0  & 0 & 2 e^{i\pi/2} \\
0 & 1 & 1 & 0  & 0 & - 2                \\
0 & 1 & 1 & 0  &  0 & 2 e^{i3\pi/2}  

\end{array}
\right)} $
\caption{The V2PM matrix for an ABCD  3-beam combiner}
\end{figure}

\subsection{The quadrature approach}
It is possible to  reformulate the mathematical model of the beam combiner in terms of products of real valued matrices and vectors. The method is based on considering the transformation of the real and imaginary parts (quadratures) of the complex field products which uniquely define the mutual coherence properties of the interfering fields. We can thus build a vector $\mathbf{J}$ of dimension $N^2$ which replaces the complex vector $\mathbf{V}$:
\begin{equation}
\mathbf{ J}=
\left(
\begin{array}{c}
\langle  A_1 A_1^* \rangle \\  
\vdots \\ 
\langle A_NA_N^*\rangle \\
\Re \langle A_1A_2^*\rangle \\
\vdots \\
 \Re\langle A_1A_N^*\rangle \\
 \Re\langle A_2A_3^*\rangle \\
 \vdots \\
 \Im\langle A_{N-1}A_N^*\rangle \\
  \Im \langle A_1A_2^*\rangle  \\
  \vdots  \\
  \Im\langle A_1A_N^*\rangle \\
   \Im\langle A_2A_3^*\rangle \\
    \vdots \\
     \Im\langle A_{N-1}A_N^*\rangle
\end{array}
\right)
\end{equation}
Here, $\Re$ is the real part, and $\Im$ the imaginary part of the complex field product.
The output of the beam combiner is now described by the real matrix $\alpha$ of dimensions MxN$^2$ :
\begin{equation}
I_\mathrm{m}=\sum_{\mathrm{n}=1}^{N^2}\alpha_{\mathrm{mn}}J_\mathrm{n}
\end{equation}
For an ideal beam combiner giving 4 quadratures for the pair m n of the input fields, the corresponding block of matrix alpha is depicted in Figure 2. The change of the base of the coherences adds non-zero elements to each block. Again, the matrix has a non vanishing null space, thus the coherences can be obtained only by means of a pseudo inversion of the $\alpha$ matrix from the output intensity measurements.

\begin{figure*}
\centering
$
\left(
\begin{array}{c c c c c c c c c}
 1  &   1 &    0 &    2 &    0 &    0 &    0 &    0 &    0\\
 1  &   1  &   0  &  -2  &   0  &   0  &   0  &   0  &   0\\
 1  &   1  &   0  &   0  &   0  &   0  &   2  &   0  &   0\\
 1   &  1  &   0  &   0  &   0  &   0  &  -2  &   0  &   0\\
 1  &   0  &   1  &   0  &   2  &   0  &   0  &   0   &  0\\
 1  &   0  &   1  &   0  &  -2  &   0  &   0  &   0  &   0\\
 1  &   0  &   1  &   0  &   0  &   0  &   0  &   2  &   0\\
 1  &   0   &  1  &   0  &   0  &   0  &   0  &  -2  &   0\\
 0  &   1  &   1  &   0  &   0  &   2  &   0  &   0  &   0\\
 0  &   1  &   1  &   0  &   0  &  -2  &   0  &   0  &   0\\
 0  &   1  &   1  &   0  &   0  &   0  &   0  &   0  &   2\\
 0  &   1  &   1  &   0  &   0  &   0  &   0  &   0  &  -2
 \end{array}
\right)$
\caption{The ABCD 3-beam combiner in the $\alpha$-matrix formalism}
\end{figure*}

\section{Discrete Beam Combiners}
\subsection{The concept}

As discussed in \citep{olDBC}, two-dimensional regular arrays of NxN coupled waveguides can be used to determine uniquely the coherence properties of N fields $\{A_{\mathrm{n}}\}$.  A conceptual scheme of the beam combination with arrays of waveguides is outlined in Figure 3 for the case N=4. In arrays of coupled waveguides, light injected into a waveguide spreads to neighboring sites 
upon propagation. By exciting several waveguides simultaneously, the field intensity of the excited waveguide mode as measured at the end of the array results from the interferometric mix of all input fields contributing in variable proportion, depending on the excited input site and length of the array. For a given configuration of input sites and array length, the discrete interference pattern 
can be related conveniently to the mutual coherences of all possible combinations of the input fields.    

\begin{figure}
\epsfig{file=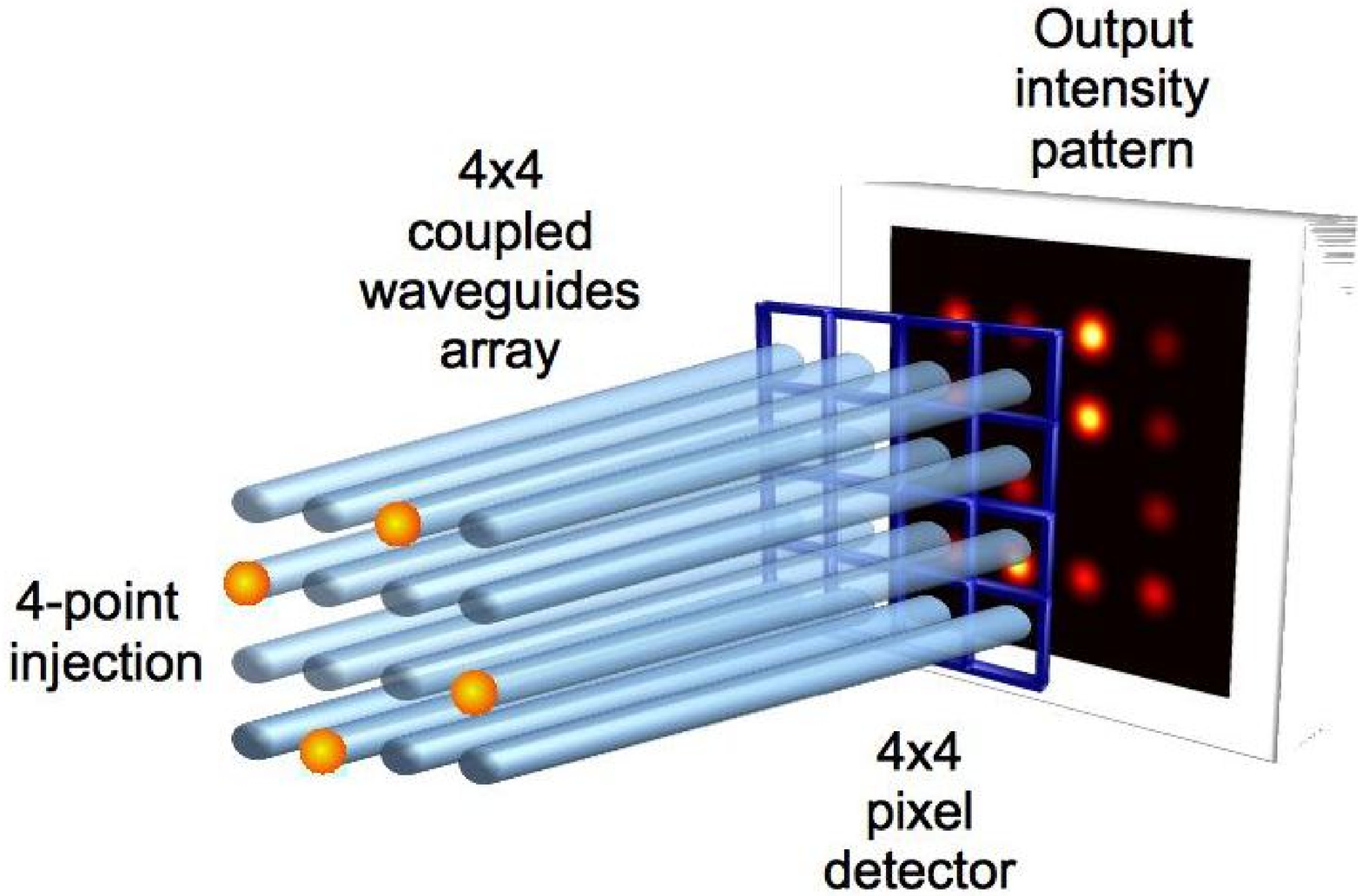, width=8cm}
\caption{A schematic view of the DBC suitable for interferometric combination of 4 telescopes. An array of 4x4=16 evanescently coupled waveguides is excited in the 4 points by the fields collected by the 4 telescopes. After suitable propagation in the array, a discrete excitation pattern of the array can be recorded (e.g. by 4x4 single pixel detectors glued to the end of the array of waveguides) which can be linearly related to the mutual coherence properties of the input fields through the inverse of the $\alpha$-matrix (see text for details).}
\end{figure}

As long as low index contrast fibers are concerned, modeling of light propagation can be carried out in terms of the coupled mode equations. 
By indicating with $E_{\mathrm{n}}$ the complex amplitude at the peak of the mode propagating in 
$\mathrm{n^{th}}$ waveguide, it is possible to describe the propagation of the fields 
along the longitudinal coordinate $z$ by means of a system of coupled differential 
equations:
\begin{equation}
i\frac{dE_{\mathrm{n}}}{dz}=\sum_{\mathrm{nm}} c_{\mathrm{nm}}E_{\mathrm{m}},
\end{equation}    
where the coupling coefficients $c_{\mathrm{nm}}$ are proportional to the 
overlap integral of the normalized complex field transverse mode profiles $u(x,y)$ of waveguides n and m. For an array of identical waveguides with core index $n_{\mathrm{core}}$, substrate index $n$ and effective propagation constant $\beta$, we can write:
\begin{equation}
\label{overlapintegral}
c_{\mathrm{nm}}=(n_{\mathrm{core}}^2-n^2)\frac{k_0^2}{2\beta}\int_\Sigma u_{\mathrm{n}}(x,y)u^{*}_{\mathrm{m}}(x,y)dxdy,
\end{equation}  
where the integral is evaluated over the cross section $\Sigma$ of the core of the fiber. Strength of the inter-waveguide coupling can be varied by means of a suitable geometric arrangement of the fibers. Solutions of equation (7) can be obtained by direct integration or by means of the supermode decomposition of the system of equations.

\subsection{Arrays of waveguides as combiners}

The $\alpha$-matrix formalism is best suited to model the output of each waveguide for a given array of coupled waveguides, because it is possible to derive an exact expression of its elements from the solution of  the coupled mode equation describing propagation of light in arrays of waveguides (Eq. 7). In fact, the peak intensity at the output of the m$^{th}$ waveguide of length z=L is given by:
\begin{eqnarray}
\label{result}
I_\mathrm{m}=\langle\left| E_\mathrm{m}\right|^2\rangle&=&\langle\left| \sum_{k=1}^{\mathrm{M}} a_{\mathrm{n,}f(\mathrm{k})} A_\mathrm{k}\right|^2\rangle \nonumber\\
&=&\sum_{j=1}^{N}\sum_{\mathrm{k}=1}^\mathrm{M} a_{\mathrm{m,}f(\mathrm{j})}a_{\mathrm{m,} f(\mathrm{k})}^* \langle A_\mathrm{j}A_\mathrm{k}^*\rangle,
\end{eqnarray}
where  and $f(\mathrm{k})$ is a function mapping k=1...N onto the sites of the NxN waveguides array where the fields $A_k$ are coupled. The coefficients $a_{\mathrm{n,} f(\mathrm{k})}$ are the mode amplitudes at the end of waveguide n when a field of unit power is injected in site f(k) at z=0.  They are  function of the array geometry and sample length and are calculated by solving 
equation (1). 
The components of the $\alpha$-matrix are then obtained straightforwardly from the following expressions:
\begin{eqnarray}
\alpha_{\mathrm{n},\mathrm{k}}&=&|a_{\mathrm{n},f(k)}|^2\quad k=1...3\\
\alpha_{\mathrm{n,p(j,k,M)}}&= & =  2\Re\{a_{\mathrm{n},f(j)}a_{\mathrm{n},f(k)}^*\}\\
\alpha_{\mathrm{n,q(j,k,M)}}&= &=  -2\Im\{a_{\mathrm{n},f(j)}a_{\mathrm{n},f(k)}^*\},
\end{eqnarray} 
where the indices p and q are defined as:
\begin{eqnarray}
\mathrm{p(j,k,M)}&=&\mathrm{ j+(k-1)\cdot(k-2)/2+M} \\
\mathrm{q(j,k,M)}&=&\mathrm{ j+(k-1)\cdot(k-2)/2+M(M+1)/2}
\end{eqnarray}
by choosing j and k according to the condition: 
\begin{equation}
\quad \mathrm{j}<\mathrm{k}\quad \mathrm{k}=2...\mathrm{M}
\end{equation}
Differently from the case of the planar beam combiner, the matrix for the DBC has in general non-zero coefficients. A sufficient condition to use of the array as an interferometric beam combiner is that the corresponding $\alpha$-matrix is invertible and well conditioned, i.e. operations involving the multiplications of the matrix or its inverse have a minimal impact on the accuracy of the result.  A practical way of gauging the performance of the combiner is to evaluate the condition number. For an invertible matrix, the condition number is defined as the ratio of the maximum to the minimal eigenvalues. In this sense, the best configuration for the combiner is the one featuring the minimal condition number. For a given array, a few input configuration can fulfill this requirement for a limited range of propagation distances. For combiners featuring a few channels only ($<10$), the best configuration can be found by means of direct numerical search. 

\subsection{Comparison with ideal combiners}

Numerical simulations of hypothetical interferometric observations (such as those presented in \citep{olDBC,SPIEDBC}) can be used to asses the advantage of one scheme over the other. The simulated observations included a photon shot-noise error source and were found to deliver performance comparable to existing beam combiners.
Here a different approach was followed and a comparison of the condition number of the DBC with that of ABCD beam combiners was performed. Even though the condition number describes the noise amplification in the worst possible scenario only, it has the advantage of being easy to compute and to represent their performance with a single number.  
In Table 1, the DBC with N ranging from 3 to 6 is compared to an equivalent ABCD combiner.

The DBC combiner model uses a square lattice geometry and  takes into account the nearest-neighbor site coupling with unitary strength in the horizontal and vertical direction, as well as a diagonal coupling to the next-nearest-neighbor sites with relative strength equal to 0.13. This model has been found to describe accurately square arrays of waveguides manufactured with the direct laser-writing technique\citep{Heinrich2009}.
The optimal input configuration and length of the DBCs  were found by direct numerical test of all possible configurations.   
Table 1 shows that the condition number of DBCs grows nearly linearly with the number of combined telescopes, with the exception of the case N=4 which features a better stability than the N=3 combiner. This fact suggests that matching the symmetry of the combination problem with that of the array could result in beam combiners of improved performance. 
Indeed, preliminary investigation, has shown that simple deformations of the array (e.g. different site separation in the vertical and horizontal direction) can sensibly reduce the condition number of the best input configuration of the DBC. Moreover, for the 6-fold-DBC, the condition number of the best configuration is smaller for hexagonal lattices. Full account of a more systematic investigation of the geometric effects will be given elsewhere. 

In all cases, the ABCD approach provides better conditioning than DBCs. This may be related to the fact that the number of interferometric measurements is larger in ABCD combiners than in DBC settings, thus resulting in higher stability. 
To prove this trend, a simple test was made by estimating the condition number for over-sized DBC arrays, i.e. arrays where the number of waveguides exceeds $N^2$. For this case, as for the case of ABCD combiners, the pseudo-inverse of the $\alpha$-matrix can be calculated. 
Table 2 resumes the result for the case of a combiner of $N$-beams on a square array of $(N+1)^2$ waveguides. The cases $N=3$ and $N=4$ are examined and show a condition number significantly lower than the classical DBC design and slightly higher than the one expected for the ABCD combiner.

\begin{table}
\begin{centering}
\mbox{
\begin{tabular}{ccccc}
\hline
N & Baselines & $\kappa_{ABCD}$ & $\kappa_{DBC}$& Configuration\\
\hline
3 & 3& 2.0 & 8.3 &\epsfig{file=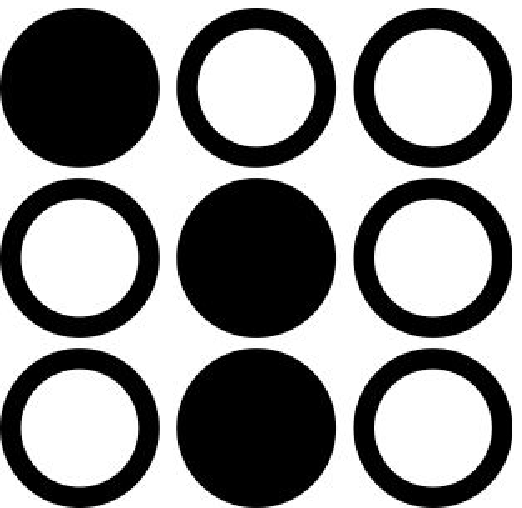,width=0.5 cm} \\
4 & 6 & 1.7 & 7.7 & \epsfig{file=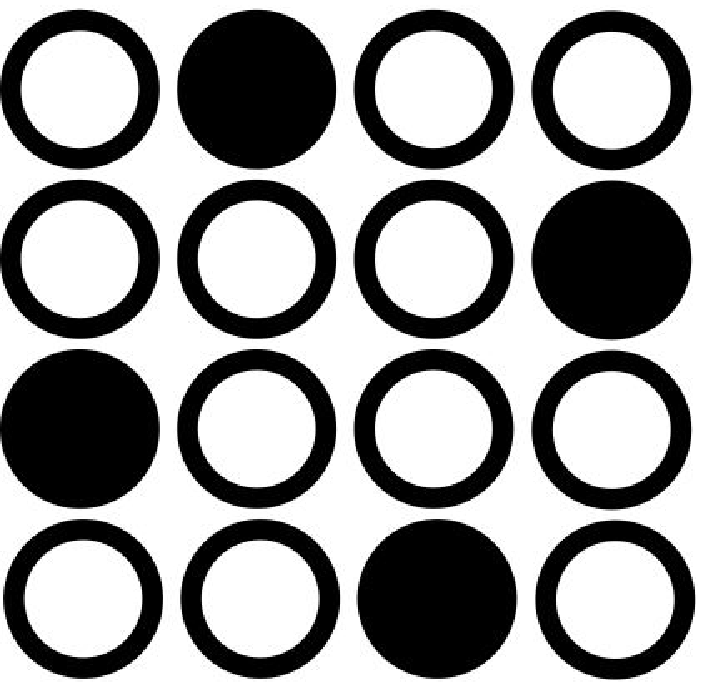, width=0.66 cm} \\
5 & 10 & 2.6 & 15.0 &\epsfig{file=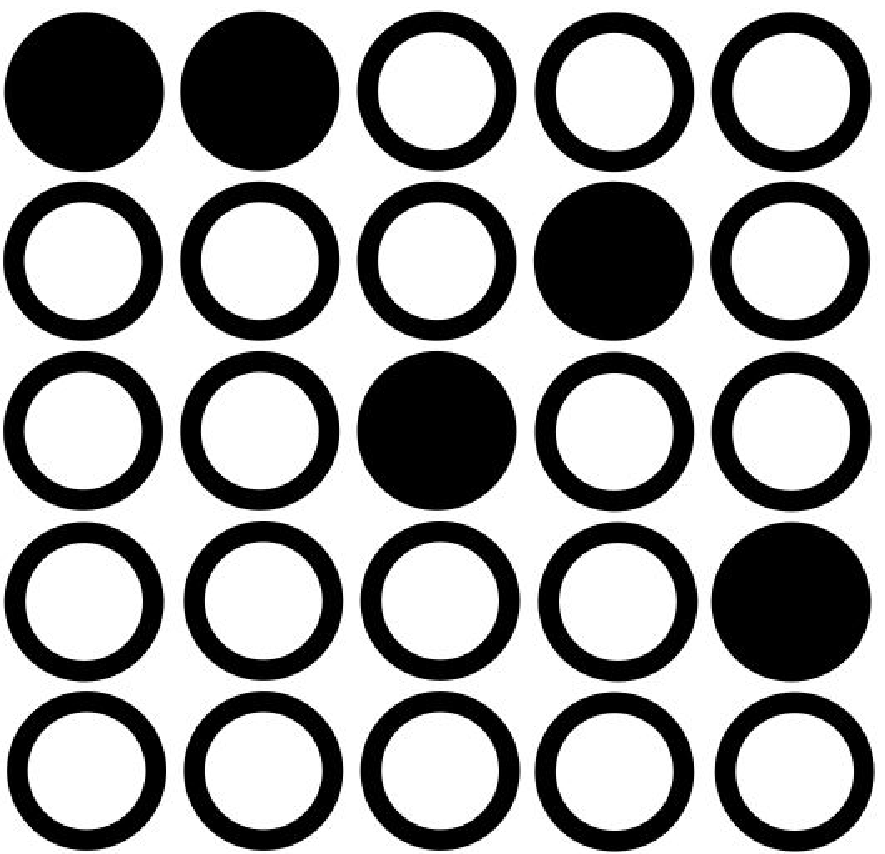, width=0.83 cm} \\
6 & 15 & 2.2 & 20.3 &\epsfig{file=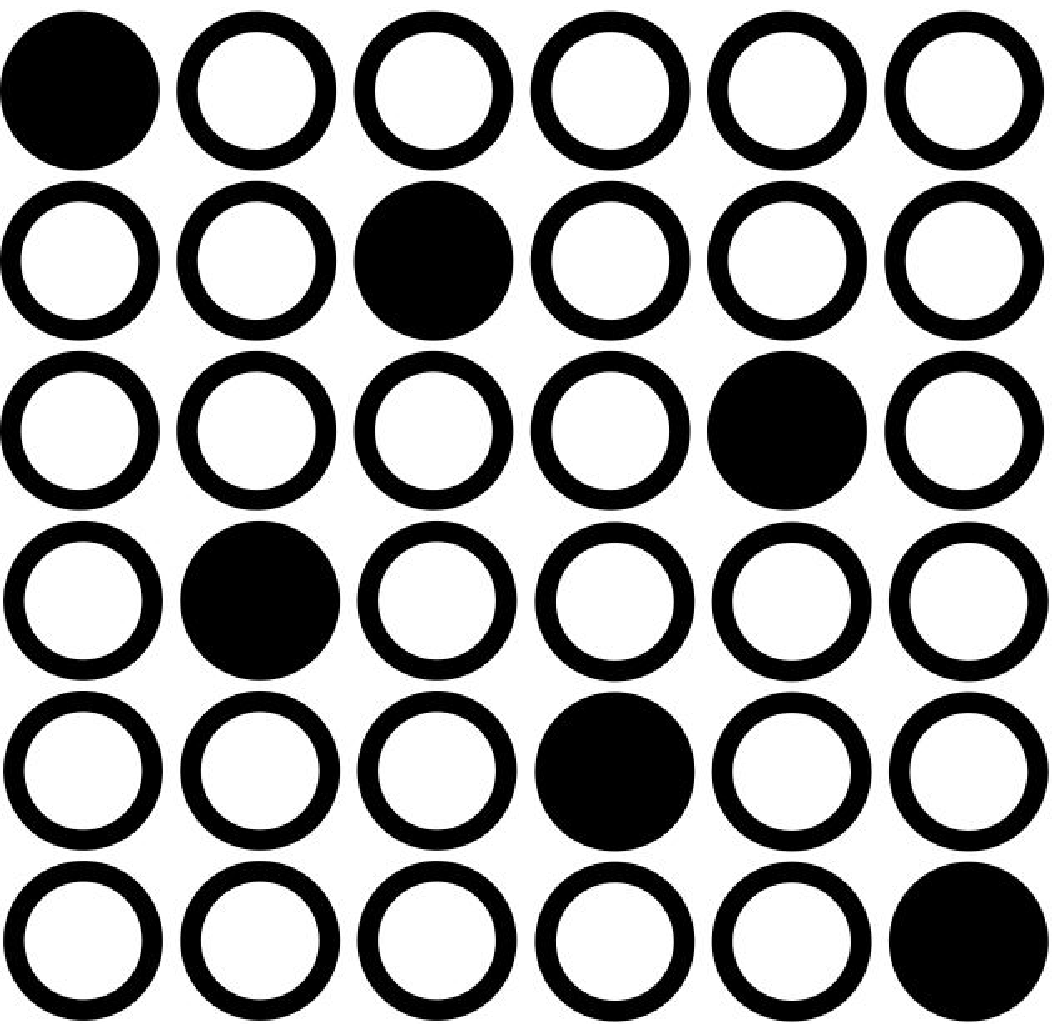, width=1 cm}  \\
\hline
\end{tabular}
}
\end{centering}
\caption{Theoretical performance of the DBC as compared to the ideal ABCD combiners. The input configuration for the best DBC is also shown (black waveguides: excitation points).}

\end{table}

\begin{table}
\begin{centering}
\mbox{
\begin{tabular}{cccc}
\hline
N & Array& $\kappa_{DBC}$ & Configuration\\
\hline
3 & 4x4 & 3.0 &\epsfig{file=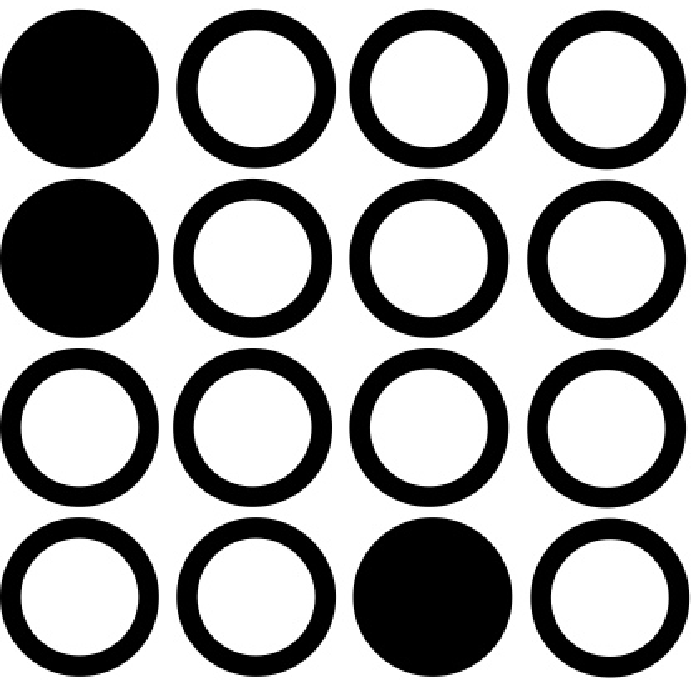,width=0.66 cm}  \\
4 & 5x5 & 4.3 &  \epsfig{file=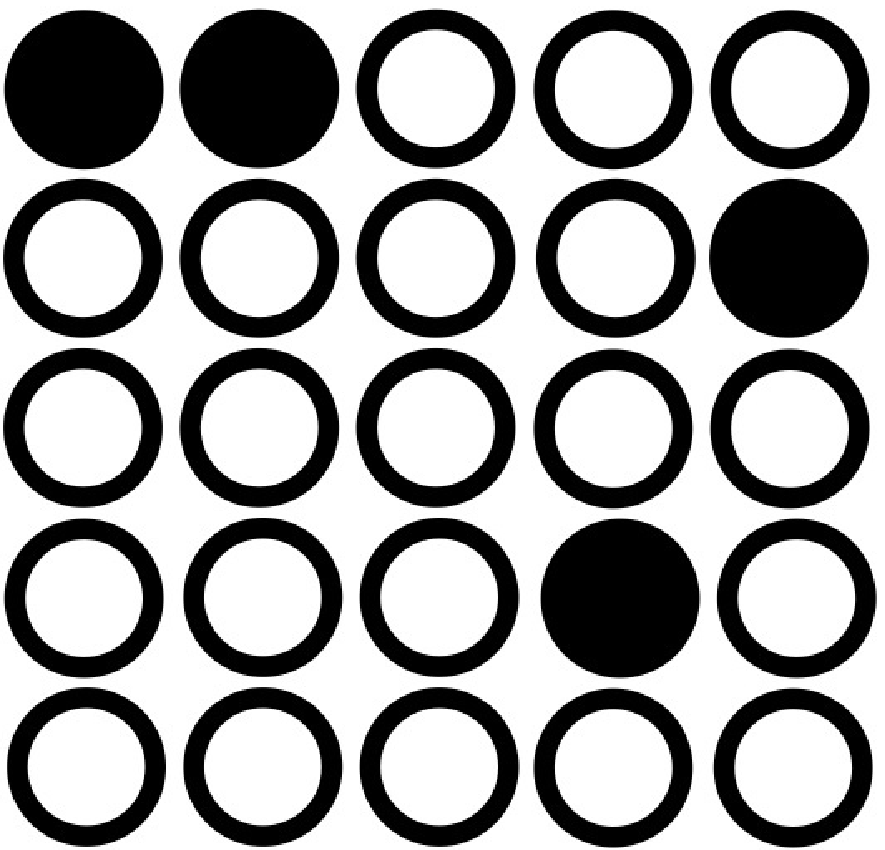, width=0.83 cm}\\ 
\hline
\end{tabular}
}
\end{centering}
\caption{Examples of oversized DBC arrays. The condition number in this case is closer to the values expected for the ABCD combiners (See Table 1).}

\end{table}

\section{Conclusions and perspectives}

A few conclusions can be drawn from the result of comparison outlined in the previous section.
A first conclusion is that ABCD combiners are better conditioned than DBCs. The price for the better performance is that the design of the component become increasingly difficult as number of baselines and combined telescopes grows.
Concerning DBCs, original arrangements for $N-$telescopes combination featuring $N^2$ arrays have condition number growing linearly with the number of combined beams, which makes them uncompetitive to ABCD (in terms of the condition number). However, DBCs featuring $M>N^2$ waveguides show a dramatic reduction of the condition number making them comparable to the ABCD combiners. 

Considering these results, it is possible to identify the potential of the DBC approach along three main lines, i.e. sensitivity, fabrication and flexibility.
Assuming that coupling losses in the structure are negligible, the DBC shares with the planar ABCD combiners a potentially higher sensitivity as compared to methods based on spatial multiaxial combiners such as in AMBER \citep{AMBER} or in pupil remapping instruments \citep{Perrin}, because the number of pixels required to read out the coherence information is much smaller. 
The ease of fabrication of DBC arrays bears the greatest potential advantage of the DBC over other methods. In fact, as opposed to intricate networks of couplers and cross-overs,  DBCs are simple arrays of waveguides which could be fabricated by rod-in-tube multicore fiber draw \citep{Roepke} or direct laser writing \cite{Itoh,Thomson09}.
The last technique entails also the potential flexibility of the DBC approach, since 
direct laser writing can be used to write photonic components on a large variety of materials including fused silica \citep{Pertsch04} and chalcogenide glasses \citep{Wong06}, thus offering the possibility to cover a wide spectral band from visible to mid infrared. Especially interesting would be the realization of such components for interferometric imaging in the mid infrared, with obvious applications to the astrophysics of sub-stellar companions or protoplanetary disks. 
Disadvantages of this technique are that the uniformity of the waveguides may critically depend on the stability of the femtosecond laser source during the writing process. Initial experiments however have shown that a non-uniformity of the coupling strength across the array of $\pm$10\% do not irremediably compromise the performance of a real DBC \citep{CLEO2011Minardi}.
Work is currently in progress to  improve the results of laboratory tests, in view of a real on sky demonstration of the capabilities of the DBC concept.

\appendix

\label{lastpage}


\begin{thebibliography}{99}
\bibitem[\protect\citeauthoryear{Benisty et al.}{2009}]{Benisty} Benisty
et al., 2009, A\&A 498, 601
\bibitem[\protect\citeauthoryear{Berger et al.}{2000}]{8combiner} Berger J. P., Benech P., SChanen I., Maury G., Malbet F., Reynaud F. 2000 Proc. SPIE 4006, 986
\bibitem[\protect\citeauthoryear{Berger et al.}{2001}]{LETI2} Berger J. P., 2001 A\&A 376, L31
\bibitem[\protect\citeauthoryear{Christodoulides et al.}{2003}]{Christodoulides} Christodoulides D., Lederer F., Silberberg Y., 2003, Nature, 424, 817 
\bibitem[\protect\citeauthoryear{Haniff et al.}{1987}]{Haniff} Haniff C.A., Mackay C.D., Titterington D.J. , et al. 1987, Nature, 323, 694
\bibitem[\protect\citeauthoryear{Heinrich et al.}{2009}]{Heinrich2009} Heinrich M.,
et al., 2009, Phys. Rev. Lett. 103, 113903.
\bibitem[\protect\citeauthoryear{Itho et al.}{2006}]{Itoh} Itoh K., Watanabe W.,  Nolte S., Schaffer C., 2006, MRS Bulletin 31, 620.
\bibitem[\protect\citeauthoryear{Kern et al.}{1996}]{LETI1} Kern, P., Malbet, F., Schanen-Duport, I., Benech, P. 1996, in AstrofibÕ 96: integrated optics for astronomical interferometry, ed. P. Kern,  F. Malbet
\bibitem[\protect\citeauthoryear{Le Bouquin et al.}{2011}]{CLEO2011LAOG} Le Bouquin, J.-B., Berger  J.-P., Lazareff, B.,  Zins, G.,  Haguenauer, P., Jocou, L., Kern P.,  Millan-Gabet, R., et al., 2011 A\&A 535, A67
\bibitem[\protect\citeauthoryear{Lacour \& Perrin}{2005}]{Perrin} Lacour S., Perrin G. , 2005 Bull. Soc. R. Sci. Liege 74, 5 
\bibitem[\protect\citeauthoryear{Lacour et al.}{2008}]{Lacour} Lacour S., et al., 2008 Proc. SPIE 7013-16 
\bibitem[\protect\citeauthoryear{Minardi \& Pertsch}{2010}]{olDBC} Minardi S., Pertsch T., 2010,
Opt. Lett., 35, 3020
\bibitem[\protect\citeauthoryear{Minardi, Neuh\"auser \& Pertsch}{2010}]{SPIEDBC} Minardi S., Neuh\"auser R., Pertsch T., 2010, Proc. SPIE, 7735, 77353P
\bibitem[\protect\citeauthoryear{Minardi et al.}{2011}]{CLEO2011Minardi} Minardi S., Chakrova N., Dreisow F., Nolte S., Pertsch T., CLEO/Europe- EQEC 2011,  22-26 May 2011, Munich, Germany, Paper JSIII1.2 THU
\bibitem[\protect\citeauthoryear{Pertsch et al.}{2004}]{Pertsch04} Pertsch T., Peschel U., Lederer F., Burgoff J., Will M., Nolte S., T\"unnermann A., 2004 Opt. Lett. 29, 468
\bibitem[\protect\citeauthoryear{Petrov et al.}{2007}]{AMBER} Petrov R. G., Malbet F., Weigelt G., Antonelli P., Beckmann U., Bresson Y., Chelli A, Dugu\'e M. et al. 2007 A\&A 464, 1
\bibitem[\protect\citeauthoryear{Quirrenbach}{2001}]{Quirrenbach} Quirrenbach A.,  2001, An. Rev. Astron. Astrophys. 39, 353
\bibitem[\protect\citeauthoryear{Rodenas et al.}{2012}]{Rodenas12} Rodenas A., Martin G., Arzeki B., Psaila N. D., Jose G., Jha A., Labadie L., Kern P., Kar A. K., Thomson R. R. 2012, Opt. Lett. 37, 392
\bibitem[\protect\citeauthoryear{R\"opke et al.}{2007}]{Roepke} R\"opcke U., Bartelt H., Unger S., Schuster K., Kobelke J., 2007, Opt. Exp. 15, 6894
\bibitem[\protect\citeauthoryear{Shao \& Staelin}{1977}]{Shao} Shao M., Staelin D. H., 1977, J. Opt. Soc. Am. 67, 81.
\bibitem[\protect\citeauthoryear{Tatulli \& LeBouquin}{2007}]{Tatulli} Tatulli E.,  LeBouquin, J.-B. 2006, MNRAS, 368, 1159
\bibitem[\protect\citeauthoryear{Thomson et al.}{2009}]{Thomson09} Thomson R. R., Kar A. K., Alligton-Smith J., 2009, Opt. Exp., 17, 1963 
\bibitem[\protect\citeauthoryear{Wong et al.}{2006}]{Wong06} Wong S., Deubel M., Perez-Willard F., John S., Ozin G. A., Wegener M., von Freymann G., 2006, Adv. Mat. 18, 265
A\&A, 196, 173
\end{thebibliography}
\end{document}